\documentclass[a4paper, amsfonts, amssymb, amsmath, reprint, showkeys, nofootinbib, twoside,longbibliography, aps]{revtex4-1}
\usepackage{lipsum}
\usepackage[english]{babel}
\usepackage[utf8]{inputenc}
\usepackage{printlen}
\usepackage[detect-all]{siunitx}
\usepackage{comment}
\usepackage[normalem]{ulem} 
\usepackage[colorinlistoftodos, color=green!40, prependcaption]{todonotes}
\usepackage{amsthm}
\usepackage{mathtools}
\usepackage{physics}
\usepackage{xcolor}
\usepackage{graphicx}
\usepackage[left=23mm,right=13mm,top=35mm,columnsep=15pt]{geometry} 
\usepackage{adjustbox}
\usepackage{placeins}
\usepackage[T1]{fontenc}
\usepackage{lipsum}
\usepackage{csquotes}
\usepackage[pdftex, pdftitle={Article}, pdfauthor={Author}]{hyperref} 

\usepackage[normalem]{ulem}

\usepackage{mathtools}

\usepackage{soul,xcolor}

\makeatletter
\newenvironment{sqcases}{%
  \matrix@check\sqcases\env@sqcases
}{%
  \endarray\right.%
}
\def\env@sqcases{%
  \let\@ifnextchar\new@ifnextchar
  \left\lbrack
  \def\arraystretch{1.2}%
  \array{@{}l@{\quad}l@{}}%
}

\usepackage{afterpage}

\begin{document}
\setstcolor{red}
\title{ Mirror-Assisted Self-Injection Locking of Laser to Whispering-Gallery-Mode Microresonator}

\author{Ramzil R. Galiev\textsuperscript{1,2}}
\email{ramzil.galiev@gmail.com}
\author{Nikita M. Kondratiev\textsuperscript{1}}
\author{Valery E. Lobanov\textsuperscript{1}}%
\author{Andrey B. Matsko\textsuperscript{3}}
\author{Igor A. Bilenko\textsuperscript{1,2}}
\affiliation{\textsuperscript{1}Russian Quantum Center, 143026 Skolkovo, Russia
}
\affiliation{\textsuperscript{2}Faculty of Physics, Lomonosov Moscow State University, 119991 Moscow, Russia}
\affiliation{\textsuperscript{3}Jet Propulsion Laboratory, California Institute of Technology, Pasadena, CA 91109-8099 USA}
\date{\today} 

\begin{abstract}
Self-injection locking is a dynamic phenomenon which provides passive stabilization of a laser emission frequency via resonant optical feedback. The stabilization coefficient depends on the level of the feedback and quality factor of the external resonant structure creating the feedback. A conventional self-injection locked laser (SIL) based on a dielectric cavity involves barely tunable resonant Rayleigh scattering (RRS). In this work we study theoretically a scheme of a SIL via a high-Q microresonator with drop-port coupled mirror,  in which optical feedback level is optimally adjusted by tuning the drop-port mirror coupling. We show that the additional reflector can improve the laser stabilization and power handling efficiency  if compared with the classic RRS-based scheme. 
\end{abstract}


\maketitle
\section{Introduction}

The self-injection locking (SIL) is an efficient method that provides passive stabilization of semiconductor lasers via resonant optical feedback from an external optical element \cite{Agrawal1984, Dahmani:87}, e.g. diffraction grating, involving  Bragg or holographic gratings in Littrow or Littman configuration \cite{Olesen_1983, Saito1982}; high finesse cavities, such as Fabry-Perot (FP) resonators \cite{Li:1989, Hollberg1988, Laurent1989} and whispering-gallery mode resonators (WGMRs) \cite{Liang:10,Liang:15,Raja2019}. A compact high-Q WGMR is an attractive optical element in a general laser stabilization scheme.  The resonator is also attractive for SIL because it does not require  modifications of commercial laser diodes or usage of fast electronic servo systems. Fueled by the increasing interest in compact spectrally pure lasers useful for precision optical measurements, the semiconductor lasers self-injection locked to WGMRs became competitive commercial products \cite{OEwaves}. Recent studies have demonstrated usefulness of high-Q optical WGMRs for stabilization of the single-frequency, multifrequency \cite{Pavlov_18np, Galiev:18} and gain-switched \cite{PhysRevApplied.15.064066} semiconductor lasers to sub-kilohertz linewidth.

Resonant Rayleigh scattering (RRS) by the internal and surface inhomogeneities of a cavity \cite{Gorodetsky:00, Kippenberg:02, Hemmerich:94, Weiss:95, Vassiliev1998, Mohageg:07, PhysRevA.83.023803, Svela2020}  is important for standard SIL lasers.  In high-Q WGMRs the internal and surface inhomogeneities are reduced at the manufacturing stage by polishing and annealing. This is done to achieve the desirable high quality factor. A small number of inhomogeneities results in reduction of the Rayleigh scattering rate of the material.  The efficiency of the RRS-based SIL also drops if the scattering rate becomes smaller than the internal attenuation of the material.

The RRS efficiency depends on the resonator loading as well as on the material attenuation \cite{Gorodetsky:00}. The maximum level of laser stabilization can be reached when the Rayleigh scattering rate is close to the loss rate resulting from the in- and out-coupling \cite{Galiev2020PhysRevApplied}. The intrinsic loss of the material has to be negligible. Achieving this condition is not a trivial task by a few reasons \cite{Chen:06, savchenkov_matsko_2019_patent, Li:12}. Firstly, the intrinsic loss is significant in the majority of materials. Secondly, balancing the attenuation rate due to the coupling and scattering leads to reduction of the resultant output power from the laser. Thirdly, keeping a resonator at a very high Q level results in an increase of the undesirable nonlinear effects in the cavity. Fourthly, the Rayleigh scattering rate is not easily tunable and optimizable in a real system.

In this work we describe a modified scheme of the SIL laser by means of a high-Q WGMR, where the optical feedback and  the laser frequency stabilization coefficients are adjusted by tuning the coupling of the mirror-assisted drop-port evanescent field coupler. In this way we avoid most of the problems associated with the RRS-based SIL scheme. We developed an analytical model for this scheme.
Our analysis shows that the maximum values of the laser frequency stabilization coefficient are approximately the same for both modified and RRS-based SIL schemes. However, in the modified scheme one can always achieve optimal SIL operation even for low RRS rates. 

To understand the physical difference between the RRS and the mirror-induced back-reflection, le us consider the following arguments. Clockwise (cw) and counter clockwise (ccw) WGMs can be excited in an ideal ring cavity. The spectrum of the modes can be described by a Lorentzian line identical for both modes. Rayleigh scattering modifies the frequency response. At the high RRS level the resonance is split due to the forward (in our case it is cw) and backward (in our case it is ccw) wave coupling. The frequency response stays the same for both the cw and ccw modes. In the scheme with the drop-port coupled mirror, considered in this paper, the situation is drastically different. The backward wave is created by the forward wave due to the reflection from the mirror, thus there is no splitting in the backward wave. The absence of such splitting simplifies the process of tuning to the optimal locking point. At the same time, the frequency spectrum of the drop port wave has a pronounced resonance splitting due to the forward and backward coupling via the mirror. In this way, the resonance splitting of the drop-port wave can be used to estimate the back-scattering rate  and the mirror coupling rate. Fundamentally, the RRS is a reciprocal effect, while the described here feedback configuration is not. The forward wave is coupled to the backward wave, while the backward wave is not coupled to the forward wave. The nonidentical spectra of the resonances observed in the add- and drop- cavity ports result from this nontrivial feature.

It was demonstrated experimentally that high-power light circulating in a monolithic microresonator leads to unwanted nonlinear effects limiting the linewidth of a self-injection locked laser \cite{Liang:15}. We found that the optimal power circulating in microresonator is approximately the same in the both classical and proposed here SIL schemes. In this paper we solved the constrained optimization problem of setting the highest possible value of the laser frequency stabilization while keeping nonlinear effects under the threshold level.

The paper is organized as follows.  The theoretical model of the WGMR coupling scheme with the  drop-port-coupled mirror is introduced in Section \ref{sec:mirror_scheme}. The theoretical model of the semiconductor laser frequency stabilization via the resonance optical feedback from the mirror-assisted WGMR is presented in Section \ref{sec:model_weak_feedback}. The appendix A contains a detailed derivation of the proposed model. The appendixes B and C consider the laser frequency stabilization under optical feedback level and nonlinear effects constrains, respectively.   

\begin{table*}[htbp]
\begin{center}
\begin{tabular}{|c|c|c|c|c|}
	\hline
	Sym. & Definition & & Sym. & Definition \\
	\hline
    $K$ & stabilization coefficient & &  $\delta_0$ & microresonator intrinsic linewidth  \\
	\hline
     $\omega$ & the system generation frequency & & $\delta_c$ & in-port coupling rate\\
    \hline
    $\omega_{d}$ & laser cavity frequency  & & $\delta_m$ & drop-port coupling rate  \\
    \hline
    $\omega_0$ & microresonator mode frequency   & & $\delta_{\Sigma}$  & total decrement: $\delta_{\Sigma} = \delta_0 + \delta_c + \delta_m$ \\
    \hline
     $\Delta \omega$ & detuning of $\omega$ from  $\omega_0$ ( $\Delta \omega = \omega - \omega_0$)  & &  $\gamma$& Rayleigh backscattering rate\\
    \hline 
    $R_m$ & mirror reflectivity coefficient ( $\tilde{R} = R_m e^{i \tilde{\psi}}$)  & & $\tau_0$ & round-trip time of microresonator  \\
    \hline
     $\tau_s$ & round-trip time between laser and microresonator  && $\tau_d$ & round-trip time of laser cavity \\
     \hline
      $\kappa_{do}$ & coupling rate between laser front-facet and microresonator & & 
      $\alpha$ & Henry factor \\
     \hline
\end{tabular}
\caption{Definition of the most important physical parameters describing the self-injection locked laser system.}
\label{tab:table}
\end{center}
\end{table*}

\section{A SIL laser with a drop-port reflector}
\label{sec:mirror_scheme}
\begin{figure}[ht]
\centering
\includegraphics[width=.5\linewidth]{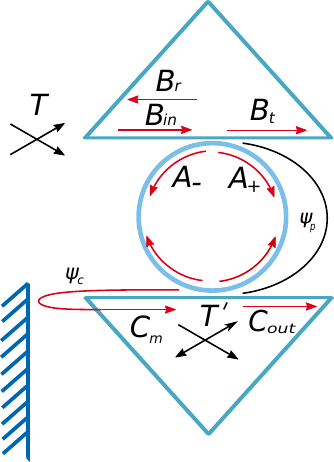}
\caption{Scheme of a SIL laser based on a ring microcavity with a mirror placed in the drop-port. The amplitudes at the \textbf{input port coupling point:}  $B_{\rm in}$ - is the amplitude of the pump; $B_{t}$ and $B_{r}$ are the amplitudes transmitted and reflected at the coupler; $A_{+}$ and $A_{-}$ are the amplitudes of the counter-rotating modes in the microresonator. The amplitudes at the \textbf{drop port coupling point:} $C_{m}$ is amplitude reflected from the mirror back to the drop-port coupler; $C_{\rm out}$ is the output amplitude from the drop-port. $T$ and $T^{'}$ are the amplitude transmittance coefficient of the in-port coupler and drop-port coupler, respectively.}
\label{fig:mirror_scheme}
\end{figure}

The scheme of the WGM-resonator-based laser stabilization setup  modified by the additional drop-port-coupled mirror is presented in Fig. \ref{fig:mirror_scheme}.  We use the quasi-geometrical approach that was introduced in \cite{Gorodetsky:99}. To analyze the impact of the  back-coupling element with a mirror on the optical feedback characteristics of a resonator we write the set of equations for the forward and backward WGM amplitudes $A_+$ and $A_-$. The equations are derived under the assumption of the ideally mode-matched single spatial mode devices. The assumption is valid for both for the laser-microresonator coupling and for the mirror-microresonator coupling. The set of equations can be presented in form
\begin{align}
\label{2_CMES_First}
    & \frac{d A_{+}}{d t} + (\delta_{\Sigma}+ i \Delta \omega) A_{+} = i \gamma A_{-} + i \frac{T}{\tau_0} B_{\rm in},\\
\label{2_CMES_Second}
    & \frac{d A_{-}}{d t} + (\delta_{\Sigma} + i \Delta \omega) A_{-} = i \gamma A_{+} + i \frac{T^{'}}{\tau_0} C_m e^{i \psi_p}, \\
\label{2_CMES_Transmittance}
    & B_t = \sqrt{1 - T^2} B_{\rm in} + i T A_{+} , \\
\label{2_CMES_Reflectance}
    & B_r = i T A_{-}, \\
\label{2_Cm}
    & C_m =i R_m T^{'} A_{+} e^{i (\psi_p +  \psi_c)}, \\
\label{2_Cout}
    & C_{\rm out} = \sqrt{1 - T^{'2}} C_m + i T^{'}A_{-} e^{-i\psi_p}.
\end{align}
Here, Eqs.~(\ref{2_CMES_First}-\ref{2_CMES_Reflectance}) are presented for the coupling point with the top prism and Eqs.~(\ref{2_Cm}-\ref{2_Cout}) are written for the coupling point with the bottom prism (see Fig. \ref{fig:mirror_scheme} and Table \ref{tab:table}). The pump wave with amplitude $B_{\rm in}$ excites the forward WGM with amplitude $A_{+}$ via the evanescent field coupler, where $T$ is the amplitude transmittance coefficient of the in-port coupler and $\tau_0$ is the resonator round-trip time. The $C_m$ is the amplitude of the wave reflected from the mirror back into the drop-port prism coupler, where $T'$ is the amplitude transmittance coefficient of the drop-port coupler. The backward wave having amplitude $A_{-}$ is excited by the forward wave $A_{+}$ due to the Rayleigh scattering on the internal and surface inhomogeneities of the resonator \cite{Gorodetsky:00, Kippenberg:02, Weiss:95,Mohageg:07, PhysRevA.83.023803, Svela2020}, where $\gamma$ is the backscattering rate. The coefficient $\delta_{\Sigma} = \delta_0 + \delta_c + \delta_m$ is the total decrement (loaded half-linewidth), where $\delta_0$ is the decrement of internal losses, the $2\delta_c=T^2/\tau_0$ is the in-port coupling rate  \cite{Gorodetsky:99} and the $2\delta_m={T'}^2/\tau_0$  is the drop-port coupling rate. $\Delta \omega = \omega  - \omega_0$ is the detuning of  the system generation frequency $\omega_d$ from the WGMR resonance frequency  $\omega_0$.  $\psi_c$ is the round-trip phase including phase shift from the mirror (see Fig. \ref{fig:mirror_scheme}). $\psi_p$ is the phase distance between two prism contacts in the forward wave direction. $R_m$ is the mirror reflectivity coefficient. Introducing  $\tilde{\psi} = \psi_c + 2 \psi_p + \pi/2$, $\tilde{R} =R_m{\rm exp}(i\tilde{\psi})$  the steady state equations for $A_{+}$ and $A_{-}$ are written in form
\begin{align}
    & (\delta_{\Sigma} + i \Delta \omega) A_{+} = i \gamma A_{-} + i \frac{2 \delta_c}{T} B_{\rm in}, \label{eq:simple_coupled_modes_1} \\
   & (\delta_{\Sigma}+ i \Delta \omega) A_{-} = i (\gamma + 2 \delta_m \tilde{R}) A_{+}, \label{eq:simple_coupled_modes_2}
\end{align}
where we took into account $\delta_m \approx T^{'2}/(2 \tau_0)$ and $\delta_c \approx T^2/(2 \tau_0)$ \cite{Gorodetsky:99}. This is a correct approximation for the case $\delta_c \tau_0 \ll 1$ and $\delta_m \tau_0 \ll 1$.  Further for the sake of simplicity we assume the optimally tuned system $\tilde R= R_m{\rm exp}(i\tilde{\psi}) = 1$. 

Introducing the reflection and transmition coefficients of the resonator as $\Gamma  = B_{\rm r}/ B_{\rm in} $  and  $T_{\rm out}  = B_{\rm t}/B_{\rm in}$, respectively, (see Fig. \ref{fig:mirror_scheme}) we derive
\begin{equation}
    \Gamma = -\frac{ 2 i \delta_c (\gamma + 2 \delta_m)}{(\delta_{\Sigma} + i \Delta \omega)^2 + \gamma  (\gamma + 2 \delta_m ) },
    \label{eq:gamma}
\end{equation}
\begin{align}
    \label{eq:T_out}
    T_{\rm out} = \frac{ (\delta_{\Sigma}-2\delta_c + i \Delta \omega)(\delta_{\Sigma} + i \Delta \omega) + \gamma  (\gamma + 2 \delta_m)}{(\delta_{\Sigma} + i \Delta \omega)^2 + \gamma  (\gamma + 2 \delta_m)},
\end{align}
\begin{align}
    \label{eq:C_out}
    |C_{\rm out}|  = |B_{\rm in}|  \left|\sqrt{\frac{\delta_m}{\delta_c}} \frac{2 \delta_c ( \delta_{\Sigma}-2\delta_m + i \Delta \omega - \gamma )}{(\delta_{\Sigma} + i \Delta \omega)^2 + \gamma  (\gamma + 2 \delta_m )}\right|,
\end{align}
where we took into account $\delta_c \tau_0 \ll 1$ and  $\delta_m \tau_0 \ll 1$ and ${\tilde{R}=1}$. The detailed derivation of Eqs.(\ref{eq:gamma}-\ref{eq:C_out}) is presented in Appendix \ref{sec:detailed_derivation_of_coupled}.
\begin{figure*}[ht]
\centering
\includegraphics[width=.9\linewidth]{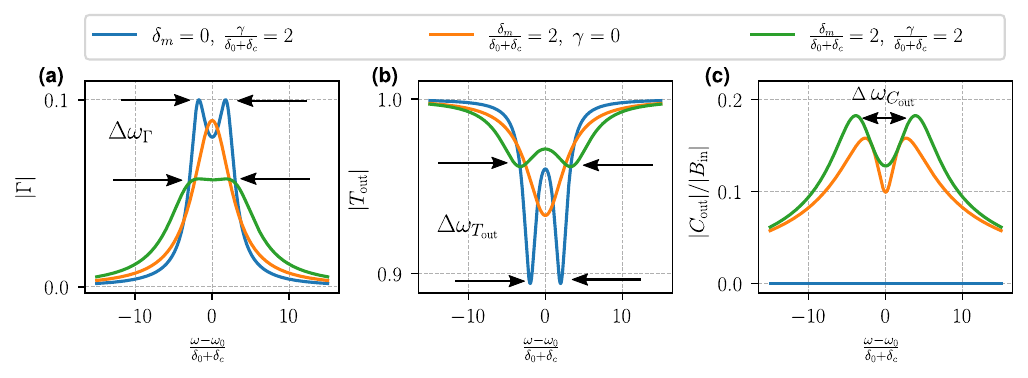}
\caption{Illustration of the spectral characteristics of the ring microcavities used in a SIL system. The resonance splitting of reflectance $\Gamma$, transmittance $T_{\rm out}$ and drop port output amplitude $C_{\rm out}$ are presented in \textbf{a, b} and \textbf{c}, respectively. The blue lines correspond to the classic SIL scheme based on RRS:$\frac{\delta_c}{\delta_0 + \delta_c} = 0.1$; $\delta_m = 0$ and $\frac{\gamma}{\delta_0 + \delta_c} = 2$. The orange lines correspond to the case of no backscattering: $\frac{\delta_c}{\delta_0 + \delta_c} = 0.1$; $\frac{\delta_m}{\delta_0 + \delta_c} = 2$ and $\frac{\gamma}{\delta_0 + \delta_c} = 0$. The green lines correspond to the general case when both the RRS and the mirror are presented: $\frac{\delta_m}{\delta_0 + \delta_c} = 2$; $\frac{\gamma}{\delta_0 + \delta_c} = 2$ and $\frac{\delta_c}{\delta_0 + \delta_c} = 0.1$.  }
\label{fig:split_resonance}
\end{figure*}
%

Coupling of the forward and backward waves causes splitting of the resonant feedback coefficient $\Gamma(\omega)$ (see Fig. \ref{fig:split_resonance}a). The splitting width is given by
\begin{equation}
    \Delta \omega_{\Gamma} \Big |_{\tilde{R}=1}=  2 \sqrt{\gamma(\gamma + 2\delta_m) - \delta_{\Sigma}^2}.
    \label{eq:gamma_splitting}
\end{equation}
The splitting is nonzero if  $\gamma  > - \delta_m + \sqrt{\delta_m^2 + \delta_{\Sigma}^2}$. Increasing the loading coefficient $\delta_m$ one can get rid of the splitting. Parameter $ \Delta \omega_{\Gamma}$ is used in the RRS-based SIL scheme to find Rayleigh scattering rate $\gamma$ \cite{Raja2019}. Similarly, one can find $\delta_m$ observing the resonance splitting of the wave leaving the drop-port wave $C_{\rm out}$
\begin{align}
\label{eq:C_out_split_omega}
        \Delta \omega_{\rm C_{\rm out}} \Big|_{\gamma = 0}^{\tilde{R}=1} = 2 \sqrt{
    (\delta_0 + \delta_c)(6\delta_m - \delta_0 - \delta_c) - \delta_m^2}.
\end{align}
The amplitude $C_{\rm out}$ has nonzero splitting width even for the case $\gamma \ll \delta_0$ (see orange lines in Fig. \ref{fig:split_resonance}). Thus, in the experiment the splitting of $C_{\rm out}$ could be used to calculate $\delta_c$ and $\delta_m$ even for the case $\gamma \ll \delta_0$. 
The amplitude $T_{\rm out}$ has zero splitting width for the case $\gamma = 0$ (see Eq. \eqref{eq:T_out_split_omega_full} and Fig. \ref{fig:split_resonance}(b)).

\section{SIL laser stabilization coefficient}
\label{sec:model_weak_feedback}

The resonant optical feedback (see Eq. \ref{eq:gamma}) to the laser results in the locking of the laser emission frequency  $\omega$ to the frequency of the resonator $\omega_0$.  A semiconductor laser model  describing weak optical feedback from a mirror to the laser was first developed by Lang-Kobayashi \cite{Lang_Kobayashi}. Models of a SIL semiconductor laser describing  weak optical feedback from a Fabry-Perot and WGMR to the laser were developed in \cite{hjelme91jqe} and \cite{Kondratiev:17}, respectively.  The models describe the tuning curve dependence of $\omega$ (the radiation frequency of the laser) on $\omega_d$ (the laser cavity frequency):
\begin{align}
    \omega- \omega_d  & =  \kappa_{do}\Im [\Gamma e^{i \omega \tau_s-i\arctan(\alpha)}], 
    \label{eq:sil_tuning_curve}
\end{align}
where $\tau_s$ is the round-trip time between the resonator and the front facet of the laser. We denoted ${\kappa_{do} = [\sqrt{1 + \alpha^2}(1-R_o^2)]/[R_o\tau_d]}$, which combines the semiconductor laser parameters: $\alpha$ is the Henry factor, $\tau_d$ is the round trip-time of the laser cavity and $R_o$ is the reflectivity of the front-facet of the laser cavity.  The coefficient $\Gamma$ provides the resonant optical feedback, which locks the radiation frequency of the laser $\omega$ to the microresonator frequency $\omega_0$. The laser can be self-injection locked to a WGMR by tuning the laser cavity frequency close to the WGMR resonance frequency ($|\omega_d - \omega_0| \sim 1.75 \delta_\Sigma\sqrt[4]{K|_{\Delta \omega = 0}}$) \cite{Shitikov2020,Kondratiev:17}, where $K|_{\Delta \omega = 0}$ is the stabilization coefficient at zero detuning.  

The resonant optical feedback results in stable lasing. The stabilization coefficient is determined by the expression
\begin{equation}
\label{eq:K_sil}
    K = \frac{d\omega_d}{d \omega}.
\end{equation}
The coefficient could be understood in the following way: if the free-running laser frequency $\omega_d$ fluctuates by the value $\delta \omega_d$, then the frequency of the self-injection locked laser $\omega$ will deviate only by $\delta \omega = \frac{\delta \omega_d}{K}$. It should be noted that this coefficient at zero detuning is also roughly equal to the locking region width, normalized to one third of the WGM linewidth.

Substituting $\Gamma$ from Eq.~\eqref{eq:gamma_full} into Eq.~\eqref{eq:sil_tuning_curve} and Eq.~\eqref{eq:K_sil}, we find 
\begin{align}
    \label{eq:full_K_derivation}
    & K =1- \Im \biggl [\frac{ 4\kappa_{do}e^{ -i\arctan(\alpha)}  \delta_c (\gamma + 2 \delta_m \tilde R)(\delta_{\Sigma} + i \Delta \omega)}{e^{-i \omega \tau_s}\left((\delta_{\Sigma} + i \Delta \omega)^2 + \gamma  (\gamma + 2 \delta_m \tilde{R} ) \right)^2} \biggr ].
\end{align}
In the case of the weak Rayleigh backscattering $\gamma \ll \delta_0$ there is no resonance splitting ($\Delta \omega_{\Gamma} = 0$ see Eq.~\eqref{eq:gamma_splitting_full}), therefore the maximum value of the stabilization coefficient is reached at $\Delta \omega = 0$. We derive for the $\tilde{\psi}$, $\Delta \omega$ and $\tau_s$ optimized stabilization coefficient of the proposed scheme 
\begin{align}
 \mathrm{arg}\max_{\tilde{\psi}, \Delta \omega, \tau_s} (K) \Big |^{|\tilde{R}|=R_m}_{\gamma \ll \delta_0} =
    \begin{cases}
        \Delta \omega = 0$, \ $\tilde{\psi} = \rm arg (\tilde R) = 0, \\
        \omega \tau_s = \arctan(\alpha) + \frac{3\pi}{2},\\ 
    \end{cases}
    \nonumber \\ 
  \max_{\tilde{\psi}, \Delta \omega, \tau_s} K \Big |_{\gamma \ll \delta_0}^{|\tilde{R}| = R_m} \approx 1+  \kappa_{do} \frac{ 4  \delta_c(\gamma + 2 \delta_m R_m)}{\delta_{\Sigma}^3}. \label{eq:simple_K_withRm}
\end{align}
The above formula can be a good approximation for ${\gamma < 0.3\delta_0}$. This approximation allows us to derive analytical expressions for the optimal parameters. 
\begin{gather}
\mathrm{arg}\max_{\delta_c, \delta_m, \tilde{\psi}, \Delta \omega, \tau_s} (K) \Big |^{|\tilde{R}|=R_m}_{\gamma \ll \delta_0} =
    \begin{cases}
   \delta_m=\delta_0-\gamma/R_m,\\ \delta_c=\delta_0-\gamma/R_m/2,\\
    \end{cases}
    \nonumber \\ 
\label{eq:global_anal}
   \max_{\delta_c, \delta_m, \tilde{\psi}, \Delta \omega, \tau_s} K \Big |_{\gamma \ll \delta_0}^{|\tilde{R}| = R_m} = 1+ \frac{16}{27}\frac{\kappa_{do}}{\delta_0} \frac{ R_m^2 }{2R_m-\gamma/\delta_0},
\end{gather}
where we took into account optimal values of  ${\Delta \omega=0}, \ \tilde{\psi} = 0$ and $\omega \tau_s = \arctan(\alpha) + \frac{3\pi}{2}$ from \eqref{eq:simple_K_withRm}. One can see that for the case $\gamma \ll \delta_0$ the maximum value of the stabilization coefficient is reached at ${R_m = 1}$ and ${\delta_m \approx \delta_c \approx \delta_0}$:
\begin{align}
   \max_{\delta_c, \delta_m, \tilde{R}, \Delta \omega, \tau_s} K \big |_{\gamma \ll \delta_0} \approx \frac{8}{27}\frac{\kappa_{do}}{\delta_0},
   \label{eq:max_for_drop_port}
\end{align}
where we take into account that $\frac{\kappa_{do}}{\delta_0} \gg 1$. 

The optimum of the stabilization coefficient of the classical SIL ($\delta_m=0$) for the case $\gamma \ll \delta_0$ is given by
\begin{align}
 &\mathrm{arg}\max_{\Delta \omega, \tau_s} K \Big |^{\delta_m = 0}_{\gamma \ll \delta_0} =
    \begin{cases}
        \Delta \omega = 0,\  \delta_c = \delta_0/2, \\
        \omega \tau_s = \arctan(\alpha) + \frac{3\pi}{2},\\ 
    \end{cases}
    \nonumber \\ 
 & \max_{\Delta \omega, \tau_s} K \Big |_{\gamma \ll \delta_0}^{\delta_m = 0} \approx 1+  \frac{16}{27} \kappa_{do}\frac{ \gamma }{\delta_0^2}. \label{eq:simple_K_withRm_low_gamma}
\end{align}
Taking into account that $\frac{\kappa_{do} \gamma }{\delta_0^2} \gg 1$ one can find
\begin{equation}
    \frac{\max_{ \Delta \omega, \tau_s} K \Big |_{\gamma \ll \delta_0}^{\delta_m = 0} }{ \max_{\delta_c, \delta_m, \tilde{R}, \Delta \omega, \tau_s} K \big |_{\gamma \ll \delta_0}  }\approx  \frac{2\gamma}{\delta_0}.
\end{equation}
Thus in the case of $\gamma \ll \delta_0$ the optimum stabilization coefficient of the drop-port mirror coupled SIL scheme is $\delta_0/(2\gamma)$ times as much as the classical SIL ($\delta_m=0$) optimum stabilization coefficient.

In \cite{Pavlov_18np} the authors measured $|\Gamma| \approx 3\times 10^{-2}$ at critical coupling ($\delta_c \approx \delta_0$) for the classic SIL scheme, which corresponds to $\gamma/\delta_0 \approx 6 \times10^{-2} $ (see Eq. \ref{eq:gamma_full}). In this way, the drop-port mirror coupled SIL scheme can enhance the stabilization coefficient  approximately by $\delta_0/(2\gamma) \approx 8$  times (or enhance the linewidth reduction by 64 times).

The optimum of drop-port scheme (see Eq.~\eqref{eq:max_for_drop_port}) corresponds to the feedback level $|\Gamma| = \frac{4}{9}$. It was shown that the strong external optical feedback $|\Gamma|$ might lead to multistability of the locked laser \cite{Lang_Kobayashi,li:1988, Vassiliev1998}, which declares another trade-off problem of setting the highest possible value of the stabilization coefficient and keeping optical feedback under the threshold level. The solution of this problem together with more details on the optimal parameters is presented in Appendix \ref{sec:optimal_regimes_and_constr}.

\begin{figure}[ht]
\centering
\includegraphics[width=1\linewidth]{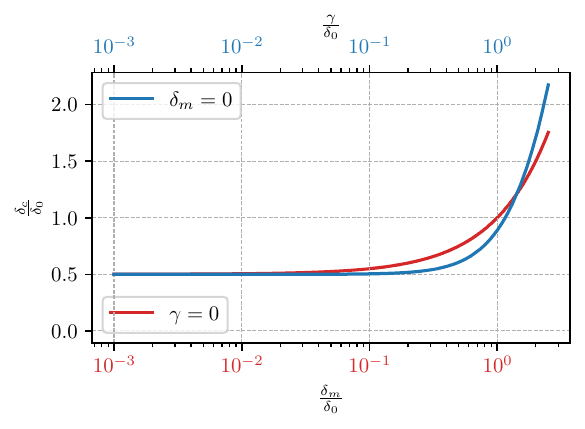}
\caption{The optimal values of $\delta_c$ for the low Rayleigh scattering case $\gamma = 0$ (blue line) and for the  classical SIL case $\delta_m = 0$ (red line). The other optimal parameters for low scattering case and for the  classical SIL case are presented in Eq.~\eqref{eq:K_classic_SIL} and Eq.~\eqref{eq:K_mirror_SIL}, respectively. 
}
\label{fig:delta_c_optimal}
\end{figure}

\begin{figure}[ht]
\centering
\includegraphics[width=1\linewidth]{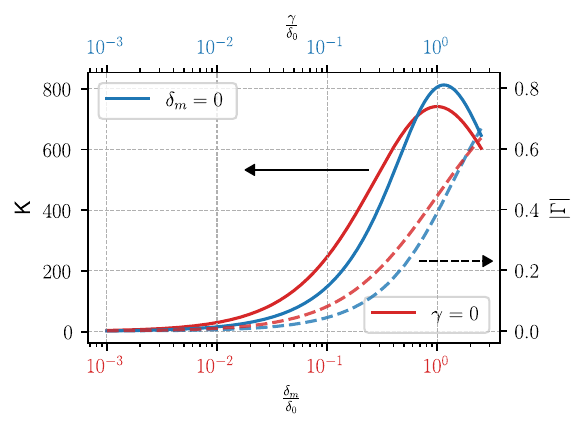}
\caption{Comparison of the optimal stabilization coefficients (\textbf{Solid lines -- left y-axis}) and the reflectance $|\Gamma|$ (\textbf{Dashed lines -- right y-axis}) for the case of negligible Rayleigh scattering (red line) and for the classical SIL case -- no drop-port mirror coupling (blue line). 
The optimal parameters for low scattering case and for the  classical SIL case are presented in Eq.~\eqref{eq:K_classic_SIL} and Eq.~\eqref{eq:K_mirror_SIL}, respectively.}
\label{fig:K_beta_and_mirror}
\end{figure}

In the considered above example we illustrated how a practical SIL can be improved using the additional reflector. It is also interesting to compare the best possible performance of the stabilization techniques when either RRS or the reflector are present (see Fig.~\ref{fig:K_beta_and_mirror}). 

\begin{align}
 &\mathrm{arg}\max_{\Delta \omega, \tau_s, \delta_c } K \Big |_{\delta_m = 0}^{\gamma \leq \delta_0 + \delta_c} =
    \begin{cases}
        \Delta \omega = 0,\ {\rm for} \ \delta_c\ {\rm see \ Fig.~\ref{fig:delta_c_optimal}}. \\
        \omega \tau_s = \arctan(\alpha) + \frac{3\pi}{2},\\ 
    \end{cases} \label{eq:K_classic_SIL}
\end{align}
where numerical calculation of the optimal $\delta_c$ and $\max_{\Delta \omega, \tau_s, \delta_c } K \Big |_{\delta_m = 0}$ are presented in Fig.~\ref{fig:delta_c_optimal} and Fig.~\ref{fig:K_beta_and_mirror}, respectively. The  $\gamma \leq \delta_0 + \delta_c$ corresponds to the $\Delta \omega \Big|_{\Gamma} = 0$ (see Eq.~\eqref{eq:gamma_splitting}), which guaranties the trivial optimum: $\Delta \omega =0$ and $\omega \tau_s = \arctan(\alpha) + \frac{3\pi}{2}$ \cite{Galiev2020PhysRevApplied}. 
\begin{align}
&\mathrm{arg}\max_{\tilde{R}, \Delta \omega, \tau_s, \delta_c} K \Big |_{\gamma = 0} =
    \begin{cases}
        \Delta \omega = 0, \ \tilde{R} = 1, \ \delta_c = \frac{\delta_0 + \delta_m}{2}, \\
        \omega \tau_s = \arctan(\alpha) + \frac{3\pi}{2},\\ 
    \end{cases} \nonumber\\
    &\max_{\tilde{R}, \Delta \omega, \tau_s, \delta_c} K \Big |_{\gamma = 0} = 1 + \frac{32}{27} \frac{\kappa_{do} \delta_m}{(\delta_0 + \delta_m)^2}
    \label{eq:K_mirror_SIL}.
\end{align}
We found that the maximal values of the stabilization coefficient for the SIL scheme with drop-port coupled mirror and for the classic SIL scheme with the optimal Rayleigh scattering reported in \cite{Galiev2020PhysRevApplied} are approximately the same (see Fig. \ref{fig:K_beta_and_mirror}). However, for the classic scheme the maximal level of laser stabilization needs precise  Rayleigh scattering rate tuning, which is not a trivial task comparing to the drop-port mirror coupling rate tuning. 

At the high level RRS rate all three resonance curves of WGMR are split due to the forward and backward wave coupling (see Eq. \ref{eq:gamma_splitting}, blue and green lines in Fig. \ref{fig:split_resonance}). In the SIL scheme with drop-port coupled mirror and low $\gamma$ the backward wave is pumped mostly by the forward wave due to reflection from the mirror, thus there is no splitting in transmittance and reflectance (see orange lines in Fig. \ref{fig:split_resonance}a,b). The absence of splitting simplifies the process of tuning to the optimal point since in this case, when changing the coupling with the prisms, there is no need to adjust the frequency detuning and phase offset.
\begin{figure}[ht]
\centering
\includegraphics[width=1\linewidth]{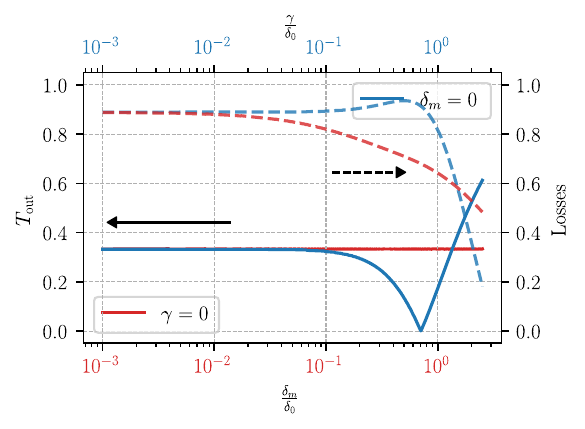}
\caption{Comparison of the transmission coefficient $T_{\rm out}$ (\textbf{Solid lines -- left y-axis}) and losses (\textbf{Dashed lines -- right y-axis}) for the case of negligible Rayleigh scattering (red line) and for the classical SIL case -- no drop-port mirror coupling (blue line). The parameters are optimized (see Eq.~\eqref{eq:K_classic_SIL} and Eq.~\eqref{eq:K_mirror_SIL}).}
\label{fig:T_and_Losses_beta_and_mirror}
\end{figure}

\begin{figure}[ht]
\centering
\includegraphics[width=1\linewidth]{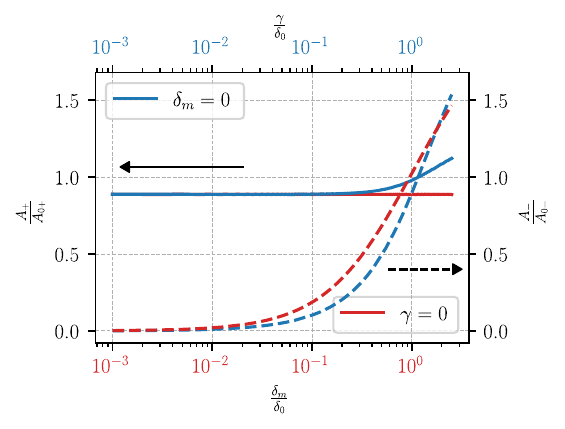}
\caption{Comparison of the normalized forward wave $\frac{A_{+}}{A_{0+}}$ (\textbf{Solid lines -- left y-axis}) and backward wave $\frac{A_{-}}{A_{0-}}$ (\textbf{Dashed lines -- right y-axis}) amplitudes for the case of negligible Rayleigh scattering (red line) and for the classical SIL case -- no drop-port mirror coupling (blue line). The parameters are chosen optimal (see Eq.~\eqref{eq:K_classic_SIL} and Eq.~\eqref{eq:K_mirror_SIL}), $A_{0\pm}$ are the amplitudes $A_{\pm}$ at $\frac{\gamma}{\delta_0} = 1.16$ and $\delta_m = 0$, which corresponds to the classical SIL optimum.}
\label{fig:A_ratios}
\end{figure}

We found that the optimal regime of the drop-port scheme is far from the critical coupling (see solid lines in Fig. \ref{fig:T_and_Losses_beta_and_mirror}) unlike the optimal regime of the classic SIL scheme. Therefore, maximization of the stabilization coefficient of the classic SIL scheme may lead to the low output power.  Introducing the radiation losses as 
\begin{align}
    {\rm Loss}=1-|\Gamma|^2-|T_{\rm out}|^2-|C_{\rm out}/B_{\rm in}|^2,
\end{align}
one can see that the losses are reduced in the scheme with reflector, compared with the classical RRS-based scheme (see dashed lines in Fig. \ref{fig:T_and_Losses_beta_and_mirror}) while the power circulating in the microresonators is approximately the same (see Fig. \ref{fig:A_ratios}). 

The high intracavity intensity can lead to unwanted nonlinear generation effects (e.g. four wave mixing or stimulated Raman scattering), which results in the transfer of the laser relative intensity noise (RIN) to the frequency noise. In previous works it was shown that unwanted nonlinear effects limit the linewidth of a self-injection locked laser \cite{Liang:15}.  The nonlinear effects inside the microresonator arise if either forward or backward wave intensity is higher than the threshold value. In the proposed scheme both forward and backward wave intensity values are close to ones of the classic SIL scheme (see Fig.\ref{fig:A_ratios}). Therefore with respect to the nonlinear effects the schemes are equivalent.

Overall, the minimization of the forward and backward wave intensity is out of the optimal parameters for the highest stabilization coefficient. This fact declares the trade-off problem of setting the highest possible value of the stabilization coefficient and keeping nonlinear effects under the threshold level, the solution of which is presented for the regime of the weak optical feedback in Appendix C.

\section{Conclusion}
\label{sec:conclusion}

We have studied the scheme of the self-injection locking of a laser via a high-Q WGMR, in which the optical feedback is adjusted by the drop-port-coupled mirror. The adjustment enables tuning of the stabilization coefficient and optimizing it for any level of Rayleigh scattering. In this way SIL scheme with a mirror solves the problem of the not optimal Rayleigh backscattering rate, which is highly suppressed in high-Q crystalline WGMRs.  
We also have noticed that the optimal regime of the proposed scheme is far from critical coupling (unlike the classic SIL scheme), which results in less radiation losses. 

\begin{acknowledgments}
The work was supported by the Russian Science Foundation (project 20-12-00344).  R.R.G. and V.E.L. acknowledge the personal support from the Foundation for the Advancement of Theoretical Physics and Mathematics "BASIS". The reported here research performed by A.M. was carried out at the Jet Propulsion Laboratory, California Institute of Technology, under a contract with the National Aeronautics and Space Administration (80NM0018D0004).
\end{acknowledgments}

\appendix
\section{}
\label{sec:detailed_derivation_of_coupled}
From equations \eqref{eq:simple_coupled_modes_1} and \eqref{eq:simple_coupled_modes_2} we obtain 
\begin{equation}
A_{+} =\frac{\delta_{\Sigma} + i \Delta \omega}{i (\gamma + 2 \delta_m \tilde{R} )} A_{-},
\label{eq:forward_backward_relation}
\end{equation}
and
\begin{equation}
    A_{-} = - \frac{B_{\rm in}}{T} \frac{2 \delta_c (\gamma + 2 \delta_m \tilde{R})}{(\delta_{\Sigma} + i \Delta \omega)^2 + \gamma  (\gamma + 2 \delta_m \tilde{R})}.
    \label{}
\end{equation}

Introducing the reflection and transmission coefficients of the resonator $\Gamma  = B_{\rm r}/ B_{\rm in} $  and  $T_{\rm out}  = B_{\rm t}/B_{\rm in}$, respectively (see Eq. \ref{2_CMES_Transmittance}-\ref{2_CMES_Reflectance}) we get:
\begin{equation}
    \Gamma = -\frac{ 2 i \delta_c (\gamma + 2 \delta_m \tilde{R})}{(\delta_{\Sigma} + i \Delta \omega)^2 + \gamma  (\gamma + 2 \delta_m \tilde{R} ) }
    \label{eq:gamma_full}
\end{equation}
and
\begin{align}
    \label{eq:T_out_append}
    T_{\rm out} = & \sqrt{1 - 2 \delta_c \tau_0} - \nonumber \\ \nonumber & - \frac{ 2  \delta_c (\delta_{\Sigma} + i \Delta \omega)}{(\delta_{\Sigma} + i \Delta \omega)^2 + \gamma  (\gamma + 2 \delta_m \tilde{R})}= \nonumber \\
                = & \sqrt{1 - 2 \delta_c \tau_0}\times\nonumber\\
    &\frac{ (\delta_{\Sigma}-2\delta_c' + i \Delta \omega)(\delta_{\Sigma} + i \Delta \omega) + \gamma  (\gamma + 2 \delta_m \tilde{R})}{(\delta_{\Sigma} + i \Delta \omega)^2 + \gamma  (\gamma + 2 \delta_m \tilde{R})},
\end{align}
where $\delta_c' = \delta_c/\sqrt{1 - 2 \delta_c \tau_0} \approx \delta_c$. The splitting width of $\Gamma$ and  $T_{\rm out}$ are given by Eq.\eqref{eq:gamma_splitting_full} and  Eq. \eqref{eq:T_out_split_omega_full}, respectively.  
\begin{equation}
    \Delta \omega_{\Gamma} \Big |_{\tilde{\psi}=0}=  2 \sqrt{\gamma(\gamma + 2R_m \delta_m) - \delta_{\Sigma}^2}.
    \label{eq:gamma_splitting_full}
\end{equation}

The critical coupling ($T_{\rm out} = 0$ in resonance) for real $\tilde R=R_m$ is reached if $\delta_c^2 = (\delta_m + \delta_0)^2 + \gamma(\gamma + 2\delta_mR_m)$, where we took into account $\delta_c \tau_0 \ll 1$.
The resonant reflection $\Gamma$ for the case of the critical coupling is given by
\begin{equation}
    \Gamma =  -i\frac{\gamma + 2\delta_m R_m}{\delta_{\Sigma}} \Big |_{2 \gamma \delta_m R_m \ll (\delta_0 +\delta_m)^2,}^{\gamma \ll 2 \delta_m R_m} \approx -i \frac{\delta_m R_m}{\delta_0 + \delta_m}. 
\end{equation}
%
The $C_{\rm out}$ is given by
\begin{align}
    \label{eq:C_out_append}
    &  C_{\rm out} =  iB_{\rm in}  \sqrt{\frac{\delta_m}{\delta_c}} e^{-i\psi_p}  \sqrt{1 - 2 \delta_m \tau_0}\times  \nonumber \\
    & \times \frac{2 \delta_c ((\delta_{\Sigma}-2\delta_m' + i \Delta \omega)\tilde{R} - \gamma' )}{(\delta_{\Sigma} + i \Delta \omega)^2 + \gamma  (\gamma + 2 \delta_m \tilde{R})},
\end{align}
where $\delta_m' = \frac{\delta_m}{\sqrt{1 - 2 \delta_m \tau_0}} \approx \delta_m$ and $\gamma' = \frac{\gamma}{\sqrt{1 - 2 \delta_m \tau_0}}\approx \gamma$.  The splitting width of $C_{\rm out}$ is given by Eq. \eqref{eq:C_out_split_omega_full}. 

The drop-port critical coupling ($C_{\rm out} = 0$) for real $\tilde R=R_m$ is given by $\delta_m = \delta_0 + \delta_c-\gamma/R_m$, which corresponds to $R_m |A_{+}| = |A_{-}|$ (see Eq. \ref{eq:forward_backward_relation}). In this way, the drop-port critical coupling is related to destructive interference between two waves in Eq. \ref{2_Cout}. Note, that the drop-port critical coupling condition is inconsistent with the input prism critical coupling, so that critical coupling for both drop-port and input port is impossible. 

The drop-port wave $C_{\rm out}$ is filtered, e.g. contains only the locked mode. To use the drop-port wave as the filtered output one might need to set the amplitude of $C_{\rm out}$ to the maximal value. The maximal amplitude of $C_{\rm out}$ is reached at
\begin{equation}
    \label{eq:delta_m_max_C_out}
    \delta_m \Big |_{\gamma = 0}^{\tilde{R} = 1} = (3 - 2 \sqrt{2}) \times (\delta_0 + \delta_c),
\end{equation}
that provides
\begin{equation}
\max_{\delta_m} C_{\rm out} \Big |_{\gamma = 0}^{\tilde{R} = 1} = \frac{i e^{-i\psi_p}}{2} \sqrt{\frac{\delta_c}{\delta_0 + \delta_c}} B_{\rm in}.
\end{equation}
\begin{widetext}
\begin{eqnarray}
    \label{eq:T_out_split_omega_full}
     & \Delta \omega_{\rm T_{\rm out}}\Big|_{\tilde{R} = 1}  = 2  \sqrt{ \Big( \sqrt{
    \gamma^2(\gamma + 2 \delta_m)^2(\delta_c - 2\delta_{\Sigma} )^2 +
    4 \gamma(\gamma + 2 \delta_m) \delta_{\Sigma}^2 (\delta_0 + \delta_m)^2 }
     -  \gamma(\gamma + 2 \delta_m) \delta_{\Sigma} \Big)/ (\delta_0 + \delta_m) - \delta_{\Sigma}^2},  \ \  \ \ \ \ \  \\
     \label{eq:C_out_split_omega_full}
    & \Delta \omega_{\rm C_{\rm out}}\Big|_{\tilde{R} = 1} = 2 \sqrt{2\sqrt{\gamma^4 + 4 \gamma^3 \delta_m  +5 \gamma^2 \delta_m^2  + 2(\delta_0 + \delta_c)( \gamma^2  + 3\gamma \delta_m  + 2 \delta_m^2)( \delta_0 + \delta_c - \gamma)}  - (\gamma + \delta_m - \delta_0 - \delta_c)^2. } \ \ \ \ \ \ \ \ \ 
\end{eqnarray}
\end{widetext}
\section{Optimal regimes under optical feedback constrain}
\label{sec:optimal_regimes_and_constr}
It was shown that the sufficiently strong external optical feedback level $|\Gamma|$ can lead to multistability of the locked laser \cite{Lang_Kobayashi,li:1988, Vassiliev1998}. Further analysis will be performed for $|\Gamma| < \rho \ll 1$, where $\rho$ is the level of the optical feedback guaranteeing the stability of a locked semiconductor laser in the setup.  The maximal value of the stabilization coefficient in the stable lasing regime is given by the following condition
\begin{gather}
    \begin{cases}
        \max_{\delta_c, \delta_m, \tilde{R}, \Delta \omega} K,\\
        |\Gamma|= \rho \ll 1, \\
    \end{cases}
\end{gather}
the solution of which can be expressed as:
\begin{gather}
    \mathrm{arg}\max_{\delta_c, \delta_m} K_0 \Big|_{|\Gamma| = \rho \ll 1} =
    \begin{cases}
        \delta_c =  \frac{\sqrt{\rho} }{2( 1 - \sqrt{\rho})} \delta_0, \\
        \delta_m = \delta_c, \\ 
    \end{cases}
    \nonumber \\ 
    \max_{\delta_0, \delta_c}   K_0 \Big|_{|\Gamma| = \rho \ll 1} =2 \frac{\kappa_{do}}{ \delta_0} ( \rho - \rho^{3/2})
     \label{eq:arg_K_max_sil},
\end{gather}
where we denoted $K_0$ as the optimal $K$ for the parameters $\tilde{R}, \Delta \omega, \tau_s$ (see Eq.~\eqref{eq:simple_K_withRm})
\begin{equation}
    K_0 = \max_{\tilde{R}, \Delta \omega, \tau_s} K \Big |_{\gamma \ll \delta_0} \approx  \kappa_{do} \frac{ 4  \delta_c(\gamma + 2 \delta_m) }{\delta_{\Sigma}^3}. 
    \label{eq:K_0_zero_detuning}
\end{equation}
By increasing the optical feedback level to $\rho = \frac{4}{9}$ at the optimally tuned the $\delta_c$ and $\delta_m$ one can increase the stabilization coefficient. 
The maximal level of the optical feedback $\rho$ at which the locking is stable depends on the laser parameters. For example, in \cite{Vassiliev1998} the ${\rho > 10^{-2}}$ provided instability of a laser and the resulting linewidth was 20 kHz. In \cite{Pavlov_18np} the laser was stable at $\rho = 3\times 10^{-2}$ and the sub-kHz linewidth was achieved. 
Stable laser frequency locking at $\rho \approx 0.5$ and significant linewidth reduction to sub-100 Hz was demonstrated in \cite{Liang2015NC}. The global maximum, according to \eqref{eq:arg_K_max_sil}, is reached at $\rho= \frac{4}{9}$.

To achieve the highest possible stabilization coefficient in the experiment, one should increase the $\delta_m$ and $\delta_c$ by decreasing the distance between the prisms and microresonator till the instability of the laser occurs. In this way the drop-port-coupled mirror  not only enhances the stabilization coefficient, but also allows to investigate the dependence of a laser stability on the optical feedback level. Stability of the locking regime of the laser by a single-mode high-Q resonator was theoretically analyzed in \cite{li:1988}, where stationary lasing condition is expressed via the laser frequency of the relaxation pulsations $\Omega_r$.
\begin{equation}
    K \delta_0 < \Omega_r.
\end{equation}

Taking into account the highest $K$ for the given $\rho$ (see Eq. \ref{eq:arg_K_max_sil}) one get $\rho - \rho^{3/2} < \frac{\Omega_r}{2\kappa_{do}} $. Hence $\rho$ is expressed only by the laser parameters. 

For the case $\gamma \ll \delta_0$ and $\tilde{R}=1$ one can estimate the enhancement of the stabilization coefficient due to a mirror coupling in feedback-constrained regime:
\begin{equation}
    \label{eq:comparision_mirror_without_mirror}
    \frac{K_0\Big |^{|\Gamma| = \rho \ll 1}_{\delta_m \neq 0}}{K_0\Big |^{|\Gamma| = \rho \ll 1}_{\delta_m = 0}} = \frac{27}{8} \frac{\delta_0}{\gamma} (\rho - \rho^{3/2}),
\end{equation}
Thus, for the case of $\frac{\delta_0}{\gamma} (\rho - \rho^{3/2}) \gg 1$ the significant increase in the stabilization coefficient can be achieved using the drop-port coupled mirror. For example, in \cite{Pavlov_18np} authors measured $|\Gamma| \approx 3\times 10^{-2}$ at critical coupling ($\delta_c \approx \delta_0$), which corresponds to $\frac{\gamma}{\delta_0} \approx 6 \times10^{-2} $ (see Eq. \ref{eq:gamma_full}). In this way, the mirror SIL scheme can enhance the stabilization coefficient  approximately 8 times (or enhance the linewidth reduction by 64 times), where took into account the optimal optical feedback level $\rho = \frac{4}{9}$. In conventional on-chip integrated microresonators Rayleigh backscattering is relatively high ($\gamma \approx \delta_0$) \cite{Raja:20}, thus there is no need to implement the SIL scheme with drop-port coupled mirror for them. 

In the experiment it might be easier to set one of the parameter $\delta_c$ or $\delta_m$, then tune the other one. For this approach the optimal parameters is given by:
\begin{gather}
    \mathrm{arg}\max_{\delta_c \ \mathrm{or}\ \delta_m} K_0 \Big |_{|\Gamma| = \rho \ll 1} \approx 
    \begin{sqcases}
        \delta_c = \frac{3}{8} \rho \delta_0 , \ \delta_m =\frac{\delta_0 + \delta_c}{2}, \\
        \delta_m = \frac{3}{8} \rho \delta_0, \ \delta_c = \frac{\delta_0 + \delta_m}{2}, \\ 
    \end{sqcases}
    \nonumber \\
       \max_{\delta_c \ \mathrm{or}\ \delta_m} K_0 \Big |_{|\Gamma| = \rho \ll 1}  \approx \frac{2}{3} \frac{\kappa_{do}}{ \delta_0} \rho.
    \label{eq:arg_K_max_sil_fixed}
\end{gather}
By tuning the drop-port coupling $\delta_m$ and the input-port coupling $\delta_c$ one can set the setup to the $\max_{\delta_m} C_{\rm out}$  (see Eq.\ref{eq:delta_m_max_C_out}). For this case we get:
\begin{gather}
    \mathrm{arg}\max_{\delta_c} K_0 \Big |^{\max_{\delta_m} C_{\rm out}}_{|\Gamma| = \rho \ll 1} =
    \begin{cases}
        \delta_m = (3 -2 \sqrt{2}) (\delta_0 + \delta_c),\\
        \delta_c = \frac{(2 - \sqrt{2})^2}{3 - 2\sqrt{2}} \delta_0 \rho, \\
    \end{cases}
    \nonumber \\
        \max_{\delta_c} K_0 \Big |^{\max_{\delta_m} C_{\rm out}}_{|\Gamma| = \rho \ll 1} =
    \frac{2 + \sqrt{2}}{4}
     \frac{\kappa_{do}}{ \delta_0} \rho,
     \label{eq:max_C_out_K_arg}
\end{gather}
where we took into account $\gamma \ll \delta_0$ and $ \tilde{R}=1$.

\begin{figure}[ht]
\centering
\includegraphics[width=1\linewidth]{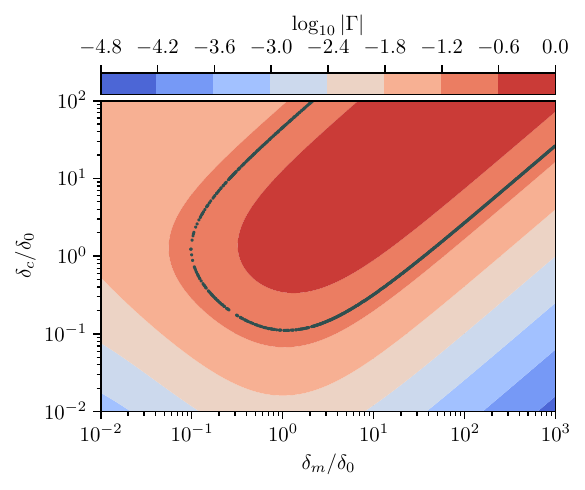}
\caption{The map of the backward reflection $|\Gamma|\big |_{\gamma =0}^{|\tilde{R}| = 1}$ (see. Eq. \ref{eq:gamma_full}) under the condition of the optimum  $\max_{\tilde{\psi}, \Delta \omega, \tau_s} K \big |_{\gamma =0}^{|\tilde{R}| = 1}$ (see Eq.~\ref{eq:simple_K_withRm}). The dark gray line corresponds to $|\Gamma| = 0.1$.}
\label{fig:Gamma_dela_c_delta_b}
\end{figure}

\begin{figure}[ht]
\centering
\includegraphics[width=1\linewidth]{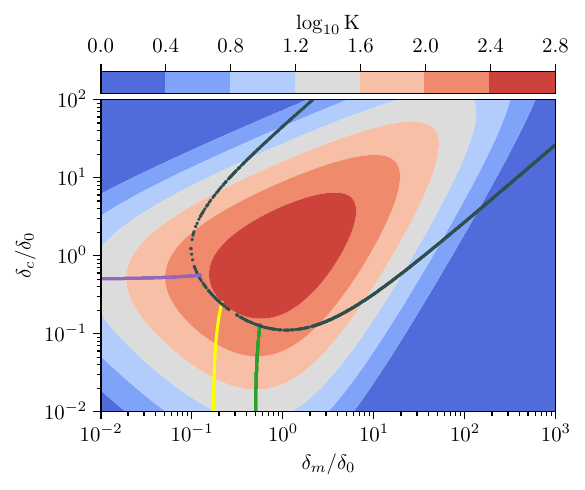}
\caption{Map of the optimal stabilization coefficient $\max_{\tilde{\psi}, \Delta \omega, \tau_s} K \big |_{\gamma =0}^{|\tilde{R}| = 1}$ (see Eq.~\ref{eq:simple_K_withRm}) . The dark gray line corresponds to $|\Gamma| = 0.1$. The purple and green lines corresponds to the $\mathrm{arg}\max (K) \Big |_{|\Gamma| \ll 1}$ (see Eq. \ref{eq:arg_K_max_sil_fixed}). The yellow line corresponds to the $\mathrm{arg}\max |C_{\rm out}|$ (see Eq. \ref{eq:max_C_out_K_arg}). }
\label{fig:K_sil_dela_c_delta_b}
\end{figure}

 The stabilization coefficient of considered regimes is close to the optimal Eq. \ref{eq:arg_K_max_sil}. The map of the stabilization coefficients is presented in Fig. \ref{fig:K_sil_dela_c_delta_b}, where the dark-gray line corresponds to the restriction $|\Gamma| = 0.1$ (see Fig. \ref{fig:Gamma_dela_c_delta_b}).

\section{Optimal regime under nonlinear effects constrain}
\label{sec:suppression_nonlinear_effects}
The high intracavity intensity can lead to unwanted nonlinear generation effects (e.g. four wave mixing or stimulated Raman scattering), which results in the transfer of the laser relative intensity noise (RIN) to the frequency noise. 

From the Eq. \ref{eq:forward_backward_relation} we get $\Big |\frac{A_{-}}{A_{+}}\Big| = \frac{ \gamma + 2 \delta_m }{\delta_{\Sigma}}$. For the case $\gamma \ll \delta_0$ (which is case for crystalline microresonators) and $\delta_m = 0$, the intracavity intensity is mostly determined by $A_{+}$ as  $|A_{-}| \ll |A_{+}|$. Thus to achieve a sufficient level of feedback $|\Gamma|$ for laser locking one need to have high amplitude of the forward wave $A_{+}$ and, consequently, strong coupling $\delta_c$. However, increase of the forward wave may lead to the unwanted nonlinear effects in high-Q microresonators. We can show that in the scheme with additional mirror varying the coupling with the mirror it is possible to tune the effective backscattering level via $\delta_m$ ($A_{-} \approx \frac{2 \delta_m }{\delta_{\Sigma}}A_{+}$) and to achieve sufficient feedback $|\Gamma|$ without significant increase of the forward wave that may prevent the occurrence of the nonlinear effects.  Further in this section the detailed analysis of the nonlinear effect suppression via the mirror coupling is presented based on the threshold of the parametric-instability, which is given by \cite{herr2014temporal,KondratievBW2019}:
\begin{align}
\label{HPO}
f=& \sqrt{\frac{\delta_c}{\delta_{\Sigma}^3}} \sqrt{\frac{6 \omega \chi_3  P_{\rm in}}{ n^4\epsilon_0 V_0}} \sqrt{\frac{n S}{n_c S_c}} > 1, 
\end{align}
where $P_{\rm in}$ is the pump power, $\chi_3$ is microresonator third order nonlinearity, $V_0$ is the mode volume, $n$ and $n_c$ are refraction indices of microresonator and coupler, $S$ and $S_c$ are beam areas in the WGM and in the coupler. The hyper-parametric oscillation and  Raman lasing have nearly identical thresholds in the WGM resonators \cite{PhysRevLett.73.2440,PhysRevA.71.033804,PhysRevLett.105.143903}.

The parameter $f$ can be decreased below the threshold level by decreasing $\delta_c$.  In this case the stabilization coefficient also decreases. In terms of keeping the high stabilization coefficient and the low value of $f$ the scheme with a mirror is more efficient than traditional SIL scheme (without a mirror).
In the new scheme, the coupling with the mirror should be taken into account and using \eqref{eq:simple_K_withRm} 
for $\gamma\ll\delta_0$ the expression for the normalized pump amplitude could be rewritten as
\begin{equation}
\label{eq:ffromK}
    f =\sqrt{ \frac{\delta_c \theta^2}{\delta_{\Sigma}^3}} \approx \sqrt{\frac{\theta^2 \max_{\tilde{\psi}, \Delta \omega, \tau_s} K \Big |_{\gamma \ll \delta_0}^{|\tilde{R}| = R_m} }{\kappa_{do}(\gamma + 2\delta_mR_m)}}, 
\end{equation}
where we denote $\theta = \sqrt{\frac{6 \omega \chi_3  P_{\rm in}}{ n^4\epsilon_0 V_0}} \sqrt{\frac{n S}{n_c S_c}}$. Analyzing this expression one can see, that for the same stabilization coefficient the drop-port mirror coupled SIL scheme has lower value of the  threshold of the parametric instability
\begin{equation}
    \frac{f|_{\delta_m \neq 0}}{f|_{\delta_m = 0}} = \sqrt{\frac{\gamma}{\gamma + 2\delta_m R_m}}.
\end{equation}
The ratio above is correct if there is no limitation on optical feedback level. In what follows we consider the system for the maximization of the stabilization coefficient under optical feedback level and nonlinear effects constrain: 
\begin{gather}
 \label{eq:arg_K_max_sil_nonlinear}
    \begin{cases}
    \max_{\delta_c, \delta_m, \tilde{R}, \Delta \omega} K , \\
         f^2 = \frac{\delta_c \theta^2 }{\delta_{\Sigma}^3} \leq 1,  \\
         |\Gamma| = \frac{4\delta_c \delta_m}{\delta_{\Sigma}^2} \leq \rho,
    \end{cases}
\end{gather}
where we took into account $\Delta \omega = 0$, $\tilde{R} = 1$ and ${\omega \tau_s = \arctan(\alpha) + \frac{3 \pi}{2}}$ from Eq. \eqref{eq:simple_K_withRm}.
Substituting \eqref{eq:arg_K_max_sil} into \eqref{eq:ffromK} we get $f^2= \frac{2\theta^2 \sqrt{\rho} (1 - \sqrt{\rho})^2}{\delta_0^2}$. So if $\frac{2\theta^2 \sqrt{\rho} (1 - \sqrt{\rho})^2}{\delta_0^2} < 1$ the optimal values of $\delta_m$ and $\delta_c$ could be set from \eqref{eq:arg_K_max_sil}, therefore for this case the  solution of \eqref{eq:arg_K_max_sil_nonlinear} is given by \eqref{eq:arg_K_max_sil}.

Further, we consider the case $\frac{2\theta^2 \sqrt{\rho} (1 - \sqrt{\rho})^2}{\delta_0^2} > 1$. In this case it is necessary to tune $\delta_m$ and $\delta_c$ out of  the optimal position of \eqref{eq:arg_K_max_sil} to satisfy $f \leq 1$. 
Combining $f^2 = \frac{\delta_c \theta^2 }{\delta_{\Sigma}^3}$ and $|\Gamma| = \frac{4\delta_c \delta_m}{\delta_{\Sigma}^2}  $ and $\delta_{\Sigma} = \delta_0 + \delta_c + \delta_m$ we get 
\begin{align}
    & \delta_m^2 -  \frac{|\Gamma|}{4f^2} \theta^2  + \frac{|\Gamma|^3}{64 f^4} \frac{\theta^4}{\delta_m^2}+ \delta_0 \delta_m   = 0.
\end{align}
Denoting $ \tilde \delta_m = \frac{2\delta_m f}{\theta \sqrt{|\Gamma|}}$ we obtain
\begin{align}
    & \tilde \delta_m^2 - 1 + \frac{|\Gamma|}{4} \frac{1}{\tilde \delta_m^2}  + \frac{2\delta_0 f}{\theta \sqrt{|\Gamma|}} \tilde \delta_m  = 0.
\end{align}
Further, we consider the case of strong nonlinearity  $\frac{2\delta_0 f}{\theta \sqrt{|\Gamma|}} \ll 1 $ and $\frac{|\Gamma|}{4}\ll\frac{\rho}{4} \ll 1$. For this case approximate solution is given by $\tilde \delta_m \approx 1$ and 
\begin{align}
    \delta_m^2  \approx  \frac{|\Gamma|}{4f^2} \theta^2.
\end{align}
Thus, we can write:
\begin{gather}
    \begin{cases}
\delta_m = \frac{\theta \sqrt{|\Gamma|}}{2 f}, \\
\delta_c   = \frac{|\Gamma| \theta}{8} \frac{\sqrt{|\Gamma|}}{f}, \\
    K\big |_{\gamma \ll \delta_0} = \frac{\kappa_{do} \sqrt{|\Gamma| f^2}}{2 \theta}. 
    \end{cases}
    \Rightarrow^{\max K }
        \begin{cases}
\delta_m = \frac{\theta \sqrt{\rho}}{2}, \\
\delta_c   = \frac{\rho \theta}{8} \sqrt{\rho}, \\
    K \big |_{\gamma \ll \delta_0} = \frac{\kappa_{do} \sqrt{\rho}}{2 \theta}, 
    \end{cases}
    \label{eq:suppress_nonlinear_sil}
\end{gather}
where to maximize $K\big |_{\gamma \ll \delta_0}$ we put $|\Gamma| = \rho$ and $f^2 = 1$. According to the system above, the optimal condition under strong nonlinearity is reached at asymmetric coupling ($\frac{\delta_c}{\delta_m} = \frac{\rho}{4}$).

\bibliography{bibliography}

\end{document}